%% file: HMIPs.tex
\documentclass[
authoryear, 12pt]{elsarticle}
\usepackage[a4paper]{geometry}
\usepackage{amssymb}

\usepackage{chngcntr}
\usepackage{caption}
\usepackage{psfrag}
\usepackage{graphicx}
\usepackage{latexsym}
\usepackage{keyval}
\usepackage{ifthen}
\usepackage{moreverb}
\usepackage{gnuplottex}
\usepackage{upgreek}
\usepackage{subfig}
 \usepackage{amsmath}
\usepackage{appendix}
\usepackage{color}
\usepackage[usenames,dvipsnames]{xcolor}
\usepackage{enumerate}
\usepackage{wrapfig}
\usepackage{titlesec}
\titlelabel{\thetitle \quad}

\usepackage{ulem}
\usepackage[section]{placeins}	
\makeatletter
\AtBeginDocument{%
	\expandafter\renewcommand\expandafter\subsection\expandafter{%
		\expandafter\@fb@secFB\subsection
	}%
}
\makeatother

\makeatletter
\AtBeginDocument{%
	\expandafter\renewcommand\expandafter\subsubsection\expandafter{%
		\expandafter\@fb@secFB\subsubsection
	}%
}
\makeatother
\usepackage[hidelinks]{hyperref}

\input{formel_preprint}

\newcommand{\avg}[1]{\langle#1\rangle}
\newcommand{\und}[1]{\underline{#1}}

\newcommand{\phitlde}{\tilde \fu}
\newcommand{\Htlde}{\tilde \feps}
\newcommand{\bH}{\bar \feps}
\newcommand{\bB}{\bar \fsigma}
\newcommand{\tcH}{\tilde {\sty \fEps}}
\newcommand{\NipFE}{N_{\rm ip}^{\rm FE}}
\newcommand{\NipHR}{N_{\rm ip}^{\rm HR}}
\newcommand{\Nmd}{N_{\rm md}}
\newcommand{\Nph}{N_{\rm ph}}
\newcommand{\Ns}{N_{\rm full}}
\newcommand{\OmegaFE}{\Omega_{\rm FE}}

\newcommand{\bBs}{\bB^s}
\newcommand{\xis}{\xi^s}
\newcommand{\seq}{\sigma_{\rm eq}}

\begin{document}
%%%%%%%%%   Your contribution (Title and authors) starts from here
\newboolean{badQuality}
\setboolean{badQuality}{true}
\setboolean{badQuality}{false}

\newlength{\smallDist} \newlength{\smallDistt}
\newlength{\smallDisttt} \newlength{\smallDistttt} 
\newlength{\vDist}

% \newcounter{zaehler} \setcounter{zaehler}{1}
% \newcounter{boxH} \setcounter{boxH}{\value{zaehler}}
% \stepcounter{zaehler} \newcounter{boxHp} \setcounter{boxHp}{\value{zaehler}}
% \stepcounter{zaehler} \newcounter{boxSig} \setcounter{boxSig}{\value{zaehler}}
% \stepcounter{zaehler} \newcounter{boxTau} \setcounter{boxTau}{\value{zaehler}}
% \stepcounter{zaehler} \newcounter{boxDivSig} \setcounter{boxDivSig}{\value{zaehler}}
% \stepcounter{zaehler} \newcounter{boxBCU} \setcounter{boxBCU}{\value{zaehler}}
% \stepcounter{zaehler} \newcounter{boxOro} \setcounter{boxOro}{\value{zaehler}}
% \stepcounter{zaehler} \newcounter{boxTauY} \setcounter{boxTauY}{\value{zaehler}}
% \stepcounter{zaehler} \newcounter{boxGND} \setcounter{boxGND}{\value{zaehler}}
% \stepcounter{zaehler} \newcounter{boxKap} \setcounter{boxKap}{\value{zaehler}}
% \stepcounter{zaehler} \newcounter{boxP} \setcounter{boxP}{\value{zaehler}}
% \stepcounter{zaehler} \newcounter{boxRhoEq} \setcounter{boxRhoEq}{\value{zaehler}}
% \stepcounter{zaehler} \newcounter{boxQEq} \setcounter{boxQEq}{\value{zaehler}}
% \stepcounter{zaehler} \newcounter{boxFQ} \setcounter{boxFQ}{\value{zaehler}}
% \stepcounter{zaehler} \newcounter{boxBCRho} \setcounter{boxBCRho}{\value{zaehler}}

\begin{frontmatter}

%% Title, authors and addresses

%% use the tnoteref command within \title for footnotes;
%% use the tnotetext command for the associated footnote;
%% use the fnref command within \author or \address for footnotes;
%% use the fntext command for the associated footnote;
%% use the corref command within \author for corresponding author footnotes;
%% use the cortext command for the associated footnote;
%% use the ead command for the email address,
%% and the form \ead[url] for the home page:
%%
%% \title{Title\tnoteref{label1}}
%% \tnotetext[label1]{}
%% \author{Name\corref{cor1}\fnref{label2}}
%% \ead{email address} 
%% \ead[url]{home page}
%% \fntext[label2]{}
%% \cortext[cor1]{}
%% \address{Address\fnref{label3}}
%% \fntext[label3]{}

% \title{{\cred Strictly Increasing Anisotropic Damage}}
%\title{Thin-film stripe shaped magnetic domain continuum theory}
\title{E3C for Computational Homogenization in Nonlinear Mechanics}
% A Thermodynamically Consistent Second Order Damage Theory With Strictly Positive Damage Growth}

%% use optional labels to link authors explicitly to addresses:
%% \author[label1,label2]{<author name>}
%% \address[label1]{<address>}
%% \address[label2]{<address>}

\author[1]{Stephan Wulfinghoff}
\author[1]{Jan Hauck}
% \author[1]{Stefanie Reese}

\address[1]{Computational Materials Science, Department of Materials Science, Kiel University, Kaiserstr.~2, 24143 Kiel, Germany\\
swu@tf.uni-kiel.de}
% \address[*]{Corresponding Author}
\begin{abstract}
In computational homogenization, a fast solution of the microscopic problem can be achieved by model order reduction in combination with hyper-reduction. Such a technique, which has recently been proposed in the context of magnetostatics, is applied to nonlinear mechanics in this work. The method is called 'Empirically Corrected Cluster Cubature' (E3C), as it combines clustering techniques with an empirical correction step to compute a novel type of integration points, which does {\it not} form a subset of the finite element integration points. The method is adopted to the challenges arising in nonlinear mechanics and is tested in plane strain for different microstructures (porous and reinforced) in dependence of the material nonlinearity. The results show that hyper-reduction errors \f{\lesssim 1}\% can be achieved with a comparably small number of integration points, which is in the order of the number of modes. A two-scale example is provided and the research code can be downloaded. 
\end{abstract}

\begin{keyword}
Computational homogenization \sep Mechanics \sep Model order reduction \sep Hyper-reduction
\end{keyword}

\end{frontmatter}
\section{Introduction}
In computational homogenization, much research has been devoted to the acceleration of the microscopic computations since the emergence of the FE\f{^2}-method \citep{smit1998prediction,miehe1999computational,kouznetsova2001approach}. For brevity, we restrict ourselves here to the discussion of reduced-order models or, more specifically, Galerkin-projection methods. These exploit that the solution of the micro-problem can often be described in a lower-dimensional subspace, which is empirically accessible using, e.g., the proper-orthogonal decomposition (POD) \citep[see][for an early two-scale method]{yvonnet2007reduced}. This shifts the computational bottleneck from the solution of the global (microscopic) equation system to the evaluation of the material law(s) at the integration points of the underlying finite element mesh. Hyper-reduction methods \citep{ryckelynck2009hyper} greatly reduce of this latter computational burden \citep[e.g.,][]{barrault2004empirical, chaturantabut2009discrete, carlberg2011efficient}, but often lack robustness \citep[as discussed by][]{van2018integration, brands2019reduced}. The empirical cubature method \citep[ECM,][]{hernandez2017dimensional} is amongst the most successful methods and identifies a small subset of the finite element integration points with optimized weights. It preserves many advantageous properties (e.g., convexity) of the underlying micro-problem and is often the method of choice  \citep[e.g.,][]{hernandez2020multiscale, lange2024monolithic}. The Empirically Corrected Cluster Cubature (E3C) method \citep{wulfinghoff2024E3c}, which is the method used here, is influenced by the ECM and related works \citep{hernandez2014high}. It departs from the assumption that the hyper-reduced integration points have to be picked from the finite element model and instead defines generalized integration points in strain space. This idea was also used in a recent work by one of the authors \citep{wulfinghoff2024statistically}, motivated by more traditional semi-analytical homogenization \citep{castaneda2016stationary}.
Attentive readers will realize that the mechanical theory presented as follows is in large parts almost identical to the magnetic derivation published previously, but with magnetic quantities replaced by mechanical ones.
% {\bf Notation.} The symmetric and skew-symmetric parts of a second-order tensor~\f{\fC} are denoted as~\f{\fsym{\fC}=(\fC+\T{\fC})/2} and~\f{\fskw{\fC}=(\fC-\T{\fC})/2}, respectively. The jump of a field quantity is written as \f{[\![\bullet]\!]}.
\section{Microscopic boundary value problem}\label{muBVP}
The problem under consideration is a microscopic boundary value problem (BVP) with an unknown position-dependent strain
\begin{equation}
 \feps(\fx)={\rm sym} \big(\grad{\fu}\big)=\bH+\Htlde(\fx)=\bH- {\rm sym} \big(\grad{\phitlde}\big).
\end{equation}
In this context, \f{\feps} consists of the macroscopic\footnote{At this stage, \f{\bH} is assumed given.} strain \f{\bH} and the strain fluctuation \f{\Htlde(\fx)}. The primary unknown is the fluctuation \f{\phitlde(\fx)} of the displacement field~\f{\fu(\fx)}. It is assumed that the fluctuation \f{\phitlde(\fx)} is periodic. The linear momentum balance in strong and weak form is expressed as follows:
\begin{equation}
 \div{\fsigma} = \fzero, \ \ \  \int\limits_\Omega \delta \Htlde  :  \fsigma \d \Omega=0,\label{weakform}.
\end{equation}
In the aforementioned equation \f{\delta \Htlde=-\grad{\delta \phitlde}} is the variation of \f{\Htlde} and \f{\Omega} denotes the domain of the periodic microstructure. For the Cauchy stress we assume a position dependent material law, which accounts for microstructure heterogeneities:
\begin{equation}
 \fsigma=\fsigma(\fx,\feps).\label{constlaw}
\end{equation}
\section{Galerkin projection}
We approximate the displacement fluctuation~\f{\phitlde} in a low-dimensional subspace:
\begin{equation}
 \phitlde(\fx) = \sum\limits_{k=1}^{\Nmd} \xi_k \tilde \fU_k(\fx).
\end{equation}
Here, \f{\tilde \fU_k(\fx)} are the given modes\footnote{The modes \f{\tilde \fU_k(\fx)} are assumed normalized and may be obtained from a representative set of finite-element simulations (snapshots) via proper orthogonal decomposition. This procedure is well documented in the cited literature and is not repeated here.} and \f{\und\xi=(\xi_1,\dots,\xi_{\Nmd})} are the unknown mode coefficients. We define associated strain modes \f{\tcH_k(\fx)={\rm sym}\left(\grad{\tilde \fU_k(\fx)}\right)}, and express the strain as follows:
\begin{equation}
 \feps(\fx) = \bH + \sum\limits_{k=1}^{\Nmd} \xi_k \tcH_k(\fx).
\end{equation}
In our case, we obtain the strain modes from a finite element (FE) model with integration points~\f{\fx^p} \f{(p=1,\dots,\NipFE)}. Therefore, we obtain for the Galerkin-projected weak form (see Eq.~\eqref{weakform}):
\begin{equation}
 \sum\limits_{k=1}^{\Nmd} \delta \xi_k \int \limits_\Omega \tcH_k  :  \fsigma \d \Omega 
 \approx \sum\limits_{k=1}^{\Nmd} \delta \xi_k \underbrace{ \left( \sum\limits_{p=1}^{\NipFE} \tcH_k(\fx^p)  :  \fsigma\big(\fx^p,\feps(\fx^p)\big) \, \OmegaFE^p\right)}_{=:R_k}=\delta \und\xi  \cdot  \und R=0. \label{discweakform}
\end{equation}
In this context, \f{\OmegaFE^p} are the integration domains of the FE integration points. 
% Further, the quantities are introduced:
% \begin{equation}
%  \tcH_k^p = \tcH(\fx^p), \ \ \ \fB^p= \fB(\fx^p,\fH^p) \ \ \ {\rm with} \ \fH^p=\bH+\sum\limits_{k=1}^{\Nmd}\xi_k \tcH_k(\fx^p).
% \end{equation}
We define the residual vector \f{\und{R}}, which must vanish since~\f{\delta \und \xi} is arbitrary:
\begin{equation}
 \und R = \begin{pmatrix} R_1\\ \vdots \\ R_{\Nmd} \end{pmatrix} = \und 0.\label{resid}
\end{equation}
The microscopic problem is solved by identifying the unknown mode coefficients~\f{\und \xi}, which satisfy Eq.~\eqref{resid} for a given \f{\bH}. Finally, the macroscopic stress \f{\bB} is computed by:
\begin{equation}
 \bB= \avg{\fsigma}= \frac{1}{\Omega} \int\limits_\Omega \fsigma \d \Omega \approx \frac{1}{\Omega} \sum\limits_{p=1}^{\NipFE} \fsigma(\fx^p,\feps(\fx^p)) \, \OmegaFE^p. \label{Bbar}
\end{equation}
% {\bf Remark:} In order to have unique strain modes, they are normalized as follows ({\cred hier nicht korrekt}):
% \begin{equation}
%  \sum\limits_{p=1}^{\NipFE} \tcH_k(\fx^p) :  \tcH_k(\fx^p) = 1, \ \ \ (k=1,\dots,\Nmd).
% \end{equation}

\section{Hyper-reduction}
\subsection{Hyper-reduction based on k-means}
This hyper-reduction method aims to reduce the computational cost associated with evaluating Eqns.~\eqref{discweakform} and \eqref{Bbar} by identifying a strongly reduced set of \f{\NipHR \ll \NipFE} integration points:
\begin{equation}
 R_k \approx \sum\limits_{q=1}^{\NipHR} \tcH_k^q  :  \fsigma^q \, \Omega^q=0,  \ \ \ \bB \approx \frac{1}{\Omega} \sum\limits_{q=1}^{\NipHR} \fsigma^q \, \Omega^q.\label{apprs}
\end{equation}
% Note that the sums now run over a significantly reduced number \f{\NipHR} of integration points and that the integration domains \f{\Omega^q} are no longer those of the finite element model and remain to be defined. The same holds for the quantities~\f{\tcH_k^q} and \f{\fsigma^q}.\\
Eq.~\eqref{apprs} now sums over a significantly reduced number of integration points. The corresponding integration weights \f{\Omega^q} differ from those of the FE model. The quantities \f{\Omega^q}, \f{\tcH_k^q} and \f{\fsigma^q} are yet to be defined. \\
As the main novelty of this paper, the hyper-reduction technique proposed in \citet{wulfinghoff2024E3c} is transferred to mechanics. The method is based on two ideas from literature:
\begin{itemize}
 \item Clustering of integration points\footnote{For related works see, e.g., \citet{liu2016self} or \citet{wulfinghoff2018model}.}
 \item Using a reduced set of integration points which preserve the original expressions in Eqns.~\eqref{discweakform} and \eqref{Bbar} as accurately as possible for Eqns.~\eqref{apprs}\f{_1} and \eqref{apprs}\f{_2} \footnote{Compare the works on ECM in the introduction.}
\end{itemize}
Initially, clusters of integration points that exhibit similar strains~\f{\feps} are identified. This is the case for two arbitrary FE integration points with indices \f{p_1} and \f{p_2} if
\begin{equation}
 \Htlde(\fx^{p_1}) = \sum\limits_{k=1}^{\Nmd} \xi_k \tcH_k(\fx^{p_1}) \approx \sum\limits_{k=1}^{\Nmd} \xi_k \tcH_k(\fx^{p_2}) = \Htlde(\fx^{p_2}).
\end{equation}
Since this is true independent of the mode coefficients~\f{\xi_k} if \f{\tcH_k(\fx^{p_1})\approx\tcH_k(\fx^{p_2})}, we cluster the FE integration points in a higher-dimensional space described by
\begin{equation}
 \tcH (\fx^p) = (\tcH_1(\fx^p), ..., \tcH_{\Nmd}(\fx^p)) \in \ffR^{d\cdot \Nmd}. \label{Htldvct}
\end{equation}
For each FE integration point \f{\fx^p} there is a counterpart \f{\tcH (\fx^p)} in this higher-dimensional space. In case of multiple phases with constitutive laws \f{\fsigma^r(\feps)}, clustering is performed separately for each phase \f{r\in (1,\dots,\Nph)}. For the clustering we use the k-means algorithm \citep{macqueen1967some} where each FE integration point is weighted by its integration domain~\f{\OmegaFE^p}. The cluster centers \f{\cC^q} \f{(q=1,\dots,\NipHR)} are computed by
\begin{equation}
 \tcH^q=\frac{1}{\Omega^q} \sum\limits_{p\in \cC^q} \tcH (\fx^p) \OmegaFE^p \ \ \ {\rm with}\ \Omega^q = \sum\limits_{p\in \cC^q} \OmegaFE^p. \label{Htldq}
\end{equation}
By this method we ensure that the strain~\f{\feps^q} at the center of a cluster~\f{\cC^q} exactly preserves the FE cluster-average (see \ref{app1} for a derivation):
\begin{equation}
 \feps^q := \bH + \sum\limits_{k=1}^{\Nmd} \xi_k \tcH^q_k = \frac{1}{\Omega^q}\sum\limits_{p\in \cC^q} \feps (\fx^p) \OmegaFE^p. \label{clustavg}
\end{equation}
Therefore, a first choice for a hyper-reduced set of \f{\NipHR} integration points is provided in terms of the cluster centers 
\begin{equation}
 \tcH^q  = (\tcH^q_1, ..., \tcH^q_{\Nmd}) \in \ffR^{d\cdot \Nmd} \ \ \ (q=1,\dots,\NipHR). \label{Htldvctq}
\end{equation}
Defining the constitutive law as 
\begin{equation}
  \fsigma^q = \fsigma^q(\feps^q).
\end{equation}
we can evaluate the hyper-reduced model in  Eq.~\eqref{apprs}. Note that \f{\fsigma^q(\feps^q)} depends on the phase the integration point belongs to (compare Eq.~\eqref{constlaw}) and that the reduced integration points can generally not be related to any of the original FE integration points.\\
% where \f{r} is the index of the phase containing cluster~\f{q}.\vspace{3mm}\\
% {\bf Remark:} If modes for the Cauchy stress fluctuation \f{\tilde \fsigma(\fx)=\fsigma(\fx)- \bB} are available in addition, the vector \eqref{Htldvct} may be extended to include also these modes with the aim to cluster finite element integration points with similar values of both, the \f{\feps}- {\it and} the \f{\fsigma}-field\footnote{This is applied in this work using five modes in the examples. However, it is expected that this does not have a significant influence.}.
\subsection{Empirical correction of the k-means integration points}
\subsubsection{Constrained cost function minimization}\label{ConstrErrMin}
In contrast to the fully integrated reduced-order model (using all \f{\NipFE} integration points), the hyper-reduced model does not exactly satisfy the Eqns.~\eqref{apprs}\f{_1} and \eqref{apprs}\f{_2}. For the empirical correction of the hyper-reduced integration points (i.e. the cluster centers) we define a cost function \f{e} of the hyper-reduced model with respect to the fully integrated reduced-order model:
\begin{align}
  e(\tcH^1, \dots, \tcH^{\NipHR}):= &\frac{1}{2} \sum\limits_{s=1}^{\Ns} \sum\limits_{k=1}^{\Nmd} \left(\frac{1}{\Omega} \sum\limits_{q=1}^{\NipHR} \tcH_k^q  :  \fsigma^q(\feps^{qs}) \, \Omega^q\right)^2 \label{errorfct1}\\
  &+ \frac{a}{2} \sum\limits_{s=1}^{\Ns} \left\| \frac{1}{\Omega}\sum\limits_{q=1}^{\NipFE} \fsigma^q(\feps^{qs}) \, \Omega^q  - \bBs \right\|^2. \label{errorfct}
\end{align}
Here, \f{a} is a user-defined weight (in this work \f{a=1}). To obtain the quantities \f{\bBs} and \f{\feps^{qs}} \f{(s=1,\dots,\Ns)} we collect \f{\Ns} simulation states with the fully integrated reduced-order model, which yield the results \f{\bBs} and  \f{\und\xis} and we define:
\begin{equation}
 \feps^{qs}(\tcH^q)=\bH^s+\sum\limits_{l=1}^{\Nmd} \xis_l \, \tcH^q_l.
\end{equation}
It is evident that the cost function~\eqref{errorfct} is similar to the works concerning the empirical cubature method (see introduction), in which it was employed to identify the reduced integration points as a subset of the FE integration points.
In this case however, minimization of the cost function \f{e} indicates that we seek for those hyper-reduced integration points (i.e., the \f{\tcH^q}) which 
\begin{enumerate}
 \item ensure that the hyper-reduced and fully integrated reduced-order models satisfy the reduced weak form of the linear momentum balance (Eq.~\eqref{discweakform}) for the same micro-states~\f{\und\xi} (expressed by Eq.~\eqref{errorfct1})
 \item yield the same macroscopic Cauchy stress \f{\bB} as the fully integrated reduced-order model (incorporated by Eq.~\eqref{errorfct}).
\end{enumerate}
We perform the minimization of the cost function~\eqref{errorfct} by enforcing the constraint 
\begin{equation}
 \sum\limits_{q=1}^{\NipHR} \tcH^q \Omega^q = \fzero
\end{equation}
through elimination of \f{\tcH^{\NipHR}}.
By this we ensure that, at any given time, the average fluctuation vanishes:
\begin{equation}
 \avg{\Htlde}= \frac{1}{\Omega}  \sum\limits_{q=1}^{\NipHR} \sum\limits_{k=1}^{\Nmd}\xi_k \tcH^q_k \Omega^q =\fzero 
 \ \ \ \Leftrightarrow \ \ \ \avg{\feps} = \bH.
\end{equation}
For the minimization, we use the Fletcher-Reeves nonlinear conjugate gradient (CG) algorithm \citep{fletcher1964function} where we take the cluster centers obtained by k-means as the starting solution. 
\subsubsection{Training procedure}\label{trainingproc}
The cost function minimization described in the previous section \ref{ConstrErrMin} is performed using linear prescribed macroscopic strain paths \f{\bar \feps(t)=\feps_0\cdot (t/T)} with \f{T=1}s and \f{\feps_0=\eps_0\fN} (\f{\|\fN\|=1}). A plane strain situation is considered from now on using Mandel-notation:
\begin{equation}
 \und{\hat \sigma}=\T{(\sigma_{11},\sigma_{22},\sqrt{2}\sigma_{12})}, \ \ \ \und{\hat \eps}=\T{(\eps_{11},\eps_{22},\sqrt{2}\eps_{12})}.
\end{equation}
The overall training procedure is divided into two stages:
\begin{enumerate}
 \item {\it Unspecific training} using \f{\sim}40 directions~\f{\und{\hat N}} being equally distributed on a unit-hemisphere (due to symmetry), applying the algorithm proposed by \citet{deserno2004generate}.
 \item {\it Feedback-controlled training} based on \f{>}100 random directions~\f{\und{\hat N}}.
\end{enumerate}
In the first stage, a fixed number of 30000 CG-iterations is applied for the cost function minimization. The second stage consists of comparing the results of the hyper-reduced and fully integrated reduced-order models and adding those five directions to the set of training directions which exhibit the largest test errors. The second stage is repeated iteratively (using 10000 CG-iterations in each iteration) until no further significant improvement is observed.\\
For a given simulation/direction \f{\und{\hat N}}, the aforementioned test error is defined by comparing the macroscopic stress responses of the hyper-reduced (HR) with the fully integrated (FI) reduced-order model:
\begin{equation}
 E=\frac{{\rm max} |\check \sigma^{\rm FI}_i-\check \sigma^{\rm HR}_i|}{{\rm max}\, \check \sigma^{\rm FI}_j - {\rm min}\, \check \sigma^{\rm FI}_k} \times 100\%,\label{errorE}
\end{equation}
where the vectors
\begin{equation}
 \und{\check \sigma}^{\rm FI} = \begin{pmatrix}
                              \langle\und{\hat \sigma}^{\rm FI}(t^*_1)\rangle\\
                              \langle \und{\hat \sigma}^{\rm FI}(t^*_2)\rangle\\
                              \vdots\\
                                \end{pmatrix}, \ \ \ 
 \und{\check \sigma}^{\rm HR} = \begin{pmatrix}
                              \langle\und{\hat \sigma}^{\rm HR}(t^*_1)\rangle\\
                              \langle \und{\hat \sigma}^{\rm HR}(t^*_2)\rangle\\
                              \vdots\\
                                \end{pmatrix}
\end{equation}
contain all macroscopic stress values at user-defined times \f{t^*_1, t^*_2, \dots}. The numerator in Eq.~\eqref{errorE} thus represents the largest stress component deviation between the two reduced-order models amongst all considered time steps of one of the \f{>}100 random directions~\f{\und{\hat N}} mentioned above.
\section{Results}
\subsection{Porous microstructure}
\subsubsection{Simulation set-up}
\begin{figure}[h]
\centering
	\hspace{-30mm}
  \begin{gnuplot}%[terminal=epslatex]
		set term epslatex size 2.5,2.4
    set xlabel '\f{\eps} [-]' offset 1,1.5,0
		set ylabel '\f{\sigma} [MPa]' offset 3, 0
		set xtics (0, 0.05)
		set ytics (0, 450)
		set key spacing 1.6
		set key at 0.047, 200
		set xrange [0:0.05]
		set yrange [0:450]
    plot 'Figs/Data/sigepsRO.dat' u 1:4 w l title '\f{p=5}', 'Figs/Data/sigepsRO.dat' u 2:4 w l  title '\f{p=10}', 'Figs/Data/sigepsRO.dat' u 3:4 w l  title '\f{p=20}'
	\end{gnuplot}	
	\hspace{-0mm}
  \psfrag{FEM}{FEM}
  \psfrag{ROM}{ROM}
  \psfrag{x1}{\f{x_1}}
  \psfrag{x2}{\f{x_2}}
  \psfrag{-315.8}{-315.8}
  \psfrag{472.5}{472.5}
  \psfrag{H}{\f{\sigma_{11}} [MPa]}
%   \psfrag{}{}
  \includegraphics[width=0.6\textwidth]{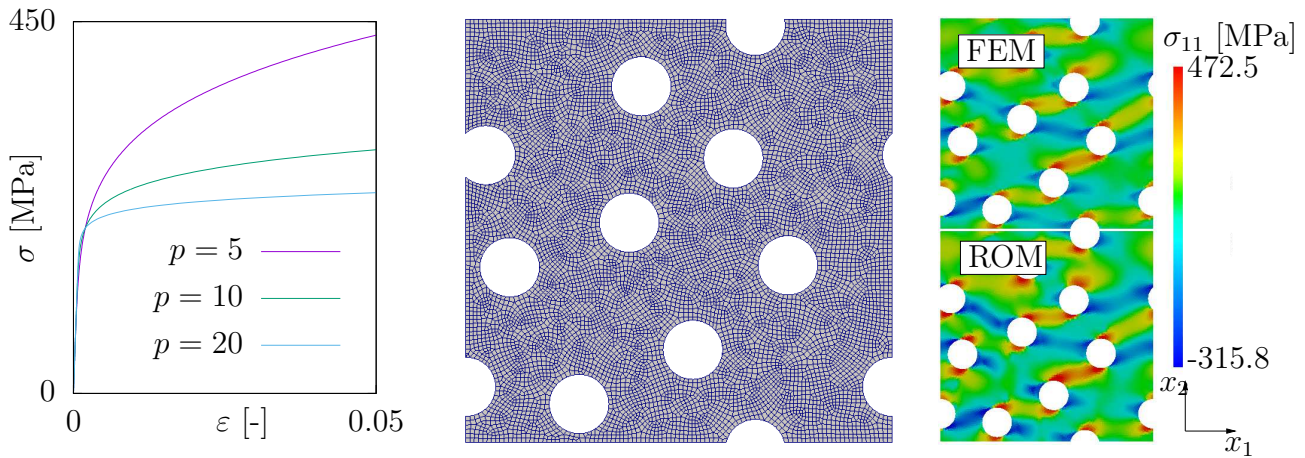}
% \caption{Left: Hyper-reduction error over number of integration points \f{\NipHR} for \f{p=5}.
\caption{Left: Ramberg-Osgood law. Center: Mesh. Right: Exemplary stress comparison (ROM reconstructed with 15 IPs)  at \f{\und{\hat \eps}=0.00707\T{(1, -1, -\sqrt{2})}} for \f{p=5}. The strain fields are not shown since they are hardly differentiable.}
\label{ROprvw}
\end{figure}

A porous microstructure with periodic boundary conditions is considered, as depicted in Fig.~\ref{ROprvw} (right). The material model applied to the matrix is the pseudo-elastoplastic Ramberg-Osgood (RO) law \citep{ramberg1943description}, which is given in detail in \ref{appRO}.  The visualization in Fig.~\ref{ROprvw} (left) shows different degrees of nonlinearity in terms of the exponent~\f{p} of the Ramberg-Osgood law, where the depicted values \f{p\in\{5,10,20\}} have been chosen to be in a typical range for most metals.
The mesh, consisting of bilinear quadrilaterals with single point integration and hourglass stabilization, is shown in Fig.~\ref{ROprvw} (center).  
% \begin{figure}[h]
% \centering
% %   \psfrag{}{}
%   \includegraphics[width=\textwidth]{Figs/meshes}
% % \caption{Left: Hyper-reduction error over number of integration points \f{\NipHR} for \f{p=5}.
% \caption{Meshes of the investigated microstructures.}
% \label{meshes}
% \end{figure}\\
For a comparably low exponent of \f{p=5}, the fully integrated reduced-order model requires \f{\Nmd=15} modes in order to yield macroscopic stress predictions, which are reasonably close (deviations in the order of 1\%) to the finite elment model\footnote{The finite element software FEAP~8.4 \citep{taylor2014feap} was used.}. Keeping in mind that the material nonlinearity is moderate, this rather high number of modes illustrates that the example of a porous microstructure is challenging, which can in part be explained by intense strain fluctuations going along with the infinite phase contrast between matrix and pores.
\subsubsection{Accuracy assessment}
\begin{figure}[h]
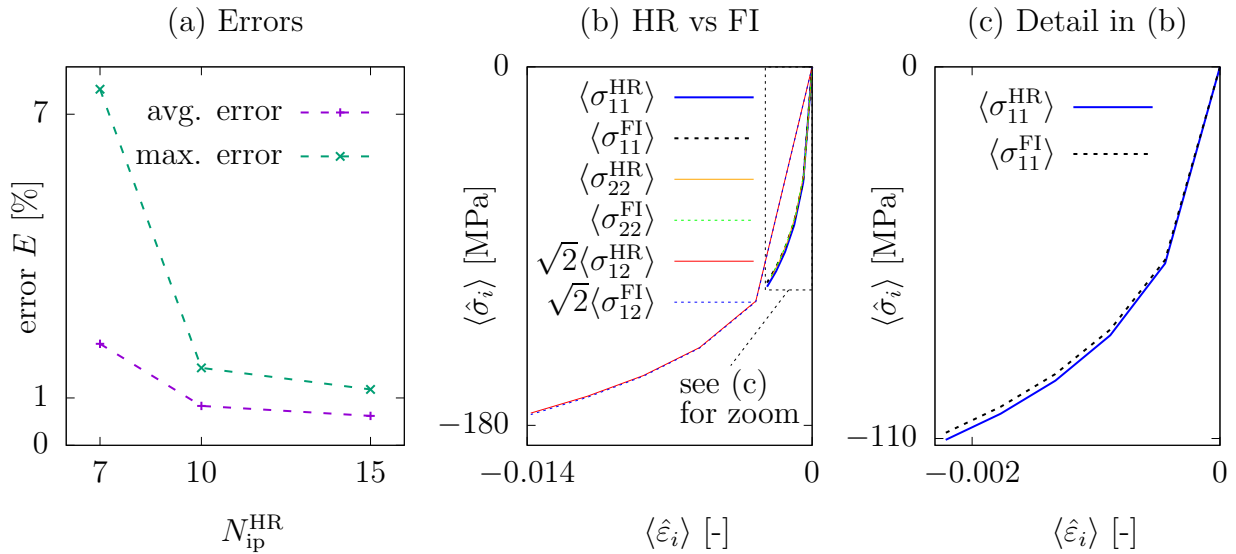

\centering
	\hspace{-15mm}
	\begin{gnuplot}%[terminal=epslatex]
		set term epslatex size 2.5,3.0 
		set title '(a) Errors'
    set xlabel '\f{\NipHR}' offset 1,-.5,0
		set ylabel 'error \f{E} [\%]' offset 1.5, 0
		set xtics (7, 10, 15)
		set ytics (0, 1,  7)
		set key spacing 1.5
		set key at 15.8, 7.5
		set xrange [6:16]
		set yrange [0:8]
    plot 'Figs/Data/unseendata_errors_pores_n5_new.dat' u 1:2 w lp title 'avg. error' dt 2 lw 3, 'Figs/Data/unseendata_errors_pores_n5_new.dat' u 1:3 w lp title 'max. error' dt 2 lw 3
	\end{gnuplot}
% 		set arrow from 6,1 to 16,1 nohead dt 2
	\hspace{-25mm}	
\begin{gnuplot}%[terminal=epslatex]
		set term epslatex size 2.5,3.0
		set title '(b) HR vs FI'
    set xlabel '\f{\langle \hat \eps_i\rangle} [-]' offset 1,-.5,0
		set ylabel '\f{\langle \hat \sigma_i \rangle} [MPa]' offset 4, 0
		set xtics (-0.014, 0)
		set ytics (-180, 0)
		set key spacing 1.4
		set key at -0.002, -5
		set xrange [-0.014:0]
		set yrange [-190:0]
		set object 1 rect from -0.0023,-112 to 0,0 fs empty dt 2
 		set arrow from -0.00115,-112 to -0.004,-150 nohead dt 2
 		set label 'see (c)' at -0.0065,-160
 		set label 'for zoom' at -0.0065,-175
    plot 'Figs/Data/sigeps_0056E3C_15IPs_n5Pores.dat' every 4 u 1:2 w l lw 3 lc rgb "blue" title '\f{\langle\sigma^{\rm HR}_{11}\rangle}', 'Figs/Data/sigeps_0056FullInt.dat' u 1:2 w l lw 3 dt 3 lc rgb "black" title '\f{\langle\sigma^{\rm FI}_{11}\rangle}', 'Figs/Data/sigeps_0056E3C_15IPs_n5Pores.dat' every 4 u 3:4 w l lc rgb "orange" title '\f{\langle\sigma^{\rm HR}_{22}\rangle}', 'Figs/Data/sigeps_0056FullInt.dat' u 3:4 w l dt 2 lc rgb "green" title '\f{\langle\sigma^{\rm FI}_{22}\rangle}', 'Figs/Data/sigeps_0056E3C_15IPs_n5Pores.dat' every 4 u 5:6 w l lc rgb "red" title '\f{\sqrt{2}\langle\sigma^{\rm HR}_{12}\rangle}', 'Figs/Data/sigeps_0056FullInt.dat' u 5:6 w l dt 2 lc rgb "blue" title '\f{\sqrt{2}\langle\sigma^{\rm FI}_{12}\rangle}'
	\end{gnuplot}
	\hspace{-25mm}	
\begin{gnuplot}%[terminal=epslatex]
		set term epslatex size 2.5,3.0
		set title '(c) Detail in (b)'
    set xlabel '\f{\langle \hat \eps_i\rangle} [-]' offset 1,-.5,0
		set ylabel '\f{\langle \hat \sigma_i \rangle} [MPa]' offset 4, 0
		set xtics (-0.002, 0)
		set ytics (-110, 0)
		set key spacing 1.6
		set key at -0.0004, -5
		set xrange [-0.0023:0]
		set yrange [-112:0]
    plot 'Figs/Data/sigeps_0056E3C_15IPs_n5Pores.dat' every 4 u 1:2 w l lw 3 lc rgb "blue" title '\f{\langle\sigma^{\rm HR}_{11}\rangle}', 'Figs/Data/sigeps_0056FullInt.dat' u 1:2 w l lw 3 dt 3 lc rgb "black" title '\f{\langle\sigma^{\rm FI}_{11}\rangle}'
	\end{gnuplot}
	\hspace{-5mm}
%     plot 'Figs/Data/20241130_08h37_light_alp0_5redIntHM2_8_2_10mds/feap/histVars.dat' u 5:($3*1000) w l title 'FEM \f{B_1}-\f{H_1}', 'Figs/Data/20241130_08h37_light_alp0_5redIntHM2_8_2_10mds/exe/histVarsRed.dat' u 4:($2*1000) w l title '10/2 IPs \f{B_1}-\f{H_1}', 'Figs/Data/20241130_08h37_light_alp0_5redIntHM2_8_2_10mds/feap/histVars.dat' u 6:($4*1000) w l title 'FEM \f{B_2}-\f{H_2}', 'Figs/Data/20241130_08h37_light_alp0_5redIntHM2_8_2_10mds/exe/histVarsRed.dat' u 5:($3*1000) w l title '10/2 IPs \f{B_2}-\f{H_2}'
\caption{(a) Hyper-reduction error \f{E} (see Eq.~\eqref{errorE}) over number of integration points \f{\NipHR} for \f{p=5}. (b) Exemplary comparison of ROM with 15 IPs (HR) vs.~full integration (FI) with the purpose to illustrate the maximum hyper-reduction error of 1.18\% amongst all 100 validation simulations. (c) Enlargement of the marked region in (b), showing the \f{\langle \sigma_{11}\rangle}-component.}
\label{errorsporesp5}
\end{figure}

The accuracy of the hyper-reduced model is evaluated by comparing its results with its fully integrated counterpart for another set of yet another 100 simulations with random strain directions~\f{\und{\hat N}}, which have not been considered during training (unseen data). The corresponding average and maximum errors are depicted in Fig.~\ref{errorsporesp5} (left) as a function of the number of integration points~\f{\NipHR}. For seven integration points, the average error of~\f{2.1\%} is already comparably low while the maximum error is still \f{7.5\%}. For 10 and 15 integration points, the observed errors reach the order of 1\% and below. The maximum error amongst the 100 validation simulations is 1.18\% for 15 integration points. The correspoding results are depicted in Fig.~\ref{errorsporesp5} (b). The maximum error occurs in the \f{\langle \sigma_{11}\rangle}-component, as illustrated in part (c) of the same figure, showing an enlargement of the region marked by the dashed rectangle in (b). The interested reader is invited to download the code ({\cblue A preliminary set of files is provided to the reviewers here: https://cloud.rz.uni-kiel.de/index.php/s/qCMTP5ZoxHZAbCf}).
\\
\subsubsection{Assessment of the computational effort}
% n=5, 10 IPs, 15 moden, Poren: doE3Ctraining: 1 min 24 s. + 2*retraining (=32 s)\\
% n=5, 30 IPs, 15 moden, Poren: doE3Ctraining: 3 min 26 s. + 7*retraining (= 6 Min 6 s)\\
% POD + clustering: 43s
% \\
The probably most important quantities to assess the computational effort of the reduced-order model are the number of modes and number of integration points needed to reach a certain accuracy. 
% Even though CPU times and related measures of interest depend on many factors like the hardware and the actual implementation, and may be considered less meaningful, some exemplary numbers are given as follows. 
The CPU-time of the hyper-reduced model with 15 integration points is on average approximately 1.1-1.3\,ms per time step (on laptop hardware\footnote{The CPU was an Intel$^{\textregistered}$ Core\f{^{\rm TM}} i7-8850H CPU @ 2.60GHz with 32 GB RAM.}), while it is in the order of 1.5\,s for the FEM-model, yielding a speed-up roughly in the order of 1200. This figure does not include the fact that, in order to reach convergence, the hyper-reduced model often needs time steps roughly half as large as the fully integrated and finite element models.\\
The training effort for the different number of integration points is summarized in Tab.~\ref{traintimes}.
\begin{table}[h]
  \centering
  \begin{tabular}{||c||c|c|c||}
  \hline
  \f{\NipHR} & 7 & 10 & 15  \\ \hline
   time & 4 min 51 s & 9 min 1 s & 18 min 24 s \\
  \hline
  \end{tabular}
  \caption{Training effort in dependence of the number of integration points~\f{\NipHR} for \f{p=5} and 15 modes (code not parallelized).}
  \label{traintimes}
\end{table}\\
The depicted figures do not include the computational effort resulting from the simulations with the fully integrated reduced-order model necessary for the training. This additional effort is in the order of 8-9 minutes for the present example.
% an hour due to the large number of finite element integration points, suggesting to replace the fully integrated model in the future by another (simple) hyper-reduced model with very high accuracy and intermediate efficiency\footnote{For example, a reduction of the currently 12217 integration points to \f{\sim}1000 integration points would already show a significant improvement.}.
% new version:\\
% n=5, 7 IPs, 15 moden, Poren: doE3Ctraining: 4 min 51 s ingesamt\\
% n=5, 10 IPs, 15 moden, Poren: doE3Ctraining: 9 min 1 s ingesamt\\
% n=5, 15 IPs, 15 moden, Poren: doE3Ctraining: 18 min 24 s ingesamt\\
\subsubsection{Influence of the nonlinearity}
If the exponent \f{p} in Eq.~\eqref{RO} is increased from 5 to 10, the hardening is decreased, as illustrated in Fig.~\ref{ROprvw} (left), leading to a stronger localization of the deformation within bands connecting the pores. This observation goes hand in hand with an increased number or modes of \f{\Nmd=25} being required to capture a reasonable accuracy in terms of macroscopic stress deviations in the order of 1\% between the finite element model and the fully integrated reduced-order model. This relatively high number of modes indicates the complexity of the combination of a porous microstructure with an increased nonlinearity (\f{p=10}). Consistently, also an increased number of \f{\NipHR=20} integration points is required to reach an average hyper-reduction error of \f{E=1.98\%} (see Eq.~\eqref{errorE}), while the maximum error is 4.07\%. For 30 and 40 integration points, the average error drops to 1.26\% and 1.0\%, respectively, while the maximum errors are given by 1.87\% and 1.59\%. It is noted that these error values depend on the training parameters. These have not been systematically optimized in this work, and more advanced training procedures may lead to lower errors.
\subsection{Reinforced composites}
\subsection{Fibre-reinforced composite}
\subsubsection{Micro-model}
\begin{figure}[h]
\centering
	\hspace{-15mm}
% 		set arrow from 6,1 to 16,1 nohead dt 2
  \psfrag{a}{(a)}
%   \psfrag{}{}
  \includegraphics[width=0.3\textwidth]{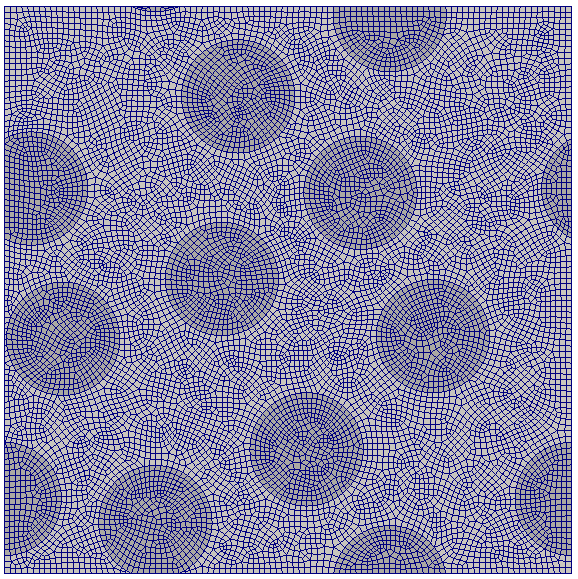}
	\hspace{-15mm}	
	\begin{gnuplot}%[terminal=epslatex]
		set term epslatex size 3.5,2.5 
		set title '(b) Max./avg. errors'
    set xlabel '\f{\NipHR}' offset 1,.5,0
		set ylabel 'error \f{E} [\%]' offset 1.5, 0
		set xtics (5, 6, 7, 8, 9)
		set ytics (0, 1, 5, 10)
		set key spacing 1.5
		set key at 10.0, 10.5
		set xrange [4.8:10.2]
		set yrange [0:11]
		set arrow from 4.8,1 to 10.2,1 nohead dt 2
    plot 'Figs/Data/unseendata_errors_Q_n5.dat' u 1:2 w lp notitle lc rgb 'red' pt 1 dt 2 lw 3, 'Figs/Data/unseendata_errors_Q_n5.dat' u 1:3 w lp title '\f{p=5}' lc rgb 'red' pt 1 dt 2 lw 3, 'Figs/Data/unseendata_errors_Q_n10.dat' u 1:2 w lp title '\f{p=10}' lc rgb 'blue' pt 3 dt 3 lw 3, 'Figs/Data/unseendata_errors_Q_n10.dat' u 1:3 w lp notitle lc rgb 'blue' pt 3 dt 3 lw 3, 'Figs/Data/unseendata_errors_Q_n20.dat' u 1:2 w lp title '\f{p=20}' lc rgb 'dark-green' pt 2  dt 4 lw 3, 'Figs/Data/unseendata_errors_Q_n20.dat' u 1:3 w lp notitle lc rgb 'dark-green' pt 2 dt 4 lw 3
	\end{gnuplot}
	\hspace{-5mm}
%     plot 'Figs/Data/20241130_08h37_light_alp0_5redIntHM2_8_2_10mds/feap/histVars.dat' u 5:($3*1000) w l title 'FEM \f{B_1}-\f{H_1}', 'Figs/Data/20241130_08h37_light_alp0_5redIntHM2_8_2_10mds/exe/histVarsRed.dat' u 4:($2*1000) w l title '10/2 IPs \f{B_1}-\f{H_1}', 'Figs/Data/20241130_08h37_light_alp0_5redIntHM2_8_2_10mds/feap/histVars.dat' u 6:($4*1000) w l title 'FEM \f{B_2}-\f{H_2}', 'Figs/Data/20241130_08h37_light_alp0_5redIntHM2_8_2_10mds/exe/histVarsRed.dat' u 5:($3*1000) w l title '10/2 IPs \f{B_2}-\f{H_2}'
\caption{(a) Mesh of the reinforced composite. (b) Maximum and average hyper-reduction errors \f{E} (see Eq.~\eqref{errorE}) over total number of integration points \f{\NipHR}. The linear-elastic fibres are represented by a single integration point.}
\label{errorsQ}
\end{figure}

Next, a fibre-reinforced composite material is investigated, the mesh of which is illustrated in Fig.~\ref{errorsQ} (left). The composite consists of elastic fibres with Young's modulus \f{E_{\rm F}=300}MPa and Poisson's ratio \f{\nu_{\rm F}=0.25} embedded in the Ramberg-Osgood matrix material described above with \f{p\in\{5,10,20\}}. In the nonlinear regime, the composite response to a given macroscopic stress is usually predominantly deviatoric in the sense that the spherical (volumetric) part of the macroscopic strain remains small. For that reason, the macroscopic strain directions used during the training procedure (see Sect.~\ref{trainingproc}) are chosen close to the deviatoric plane, such that the macroscopic stress components remain within a reasonable range \f{\lesssim 400}MPa.\\
Depending on the value of the exponent \f{p}, 11-14 modes turn out to be sufficient to decrease the error of the fully integrated reduced order model to a negligible value in comparison with the finite element model. Figure \ref{errorsQ} (right) summarizes the hyper-reduction error \f{E} for 100 validation simulations for exponents \f{p\in\{5,10,20\}}. Even in the most nonlinear case \f{p=20}, less then 10 integration points are sufficient to reach errors \f{\lesssim}1\%. In all cases, the training effort\footnote{Again, this excludes the generation of the FEM-data and the simulations using the fully integrated model.} was 5-10 minutes. 
% p=5: Sehr wenige (7) snapshots benutzt, daher mehr moden benoetigt\\
% p=5, Q, 6+1 IPs, 12 mds; doE3Ctraining: 3 min 25s; retraining: 38s+33s+49s+42s\\
% p=5, Q, 5+1 IPs, 12 mds; doE3Ctraining: 2 min 45s; retraining: 35s+37s+38s+42s+46s+49s+53s\\
% p=5, Q, 4+1 IPs, 12 mds; doE3Ctraining: 3 min 10s; retraining: 26s\\
% p=10, Q, 5+1 IPs, 11 mds; doE3Ctraining: 2 min 55 s; retraining: 29s \\
% p=10, Q, 6+1 IPs, 11 mds; doE3Ctraining: 2 min 48 s; retraining: 35 s \\
% p=10, Q, 7+1 IPs, 11 mds; doE3Ctraining: 3 min 04 s; retraining: 39s + 38s + 37s + 42s \\
% p=20, Q, 6+1 IPs, 14 mds; doE3Ctraining: 2 min 53 s; retraining: 33s + 47s \\
% p=20, Q, 7+1 IPs, 14 mds; doE3Ctraining: 4 min 28 s; retraining: 50s + s \\
% p=20, Q, 8+1 IPs, 14 mds; doE3Ctraining: 5 min 6 s; retraining: 39s + 40 s \\
\subsubsection{Two-scale simulation}
\begin{figure}[h]
\centering
	\hspace{-5mm}
% 		set arrow from 6,1 to 16,1 nohead dt 2
  \psfrag{sxx}{\f{\bar \sigma_{11}} [MPa]}
  \psfrag{-630}{-460}
  \psfrag{630}{460}
%   \psfrag{}{}
  \includegraphics[width=0.5\textwidth]{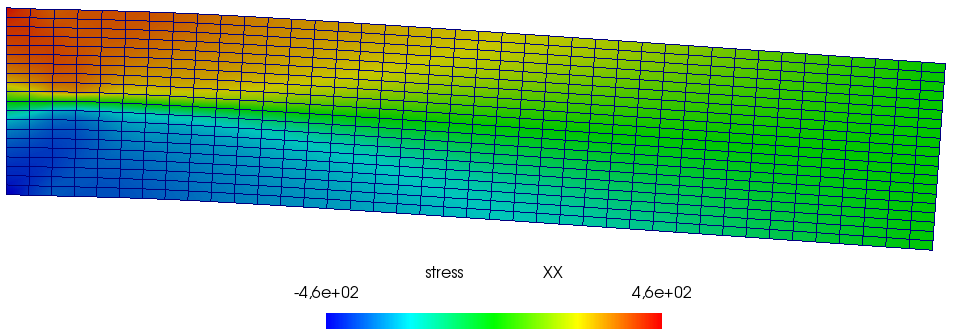}
% 	\vspace{15mm}
	\hspace{-15mm}	
% 		set title '(b) Max./avg. errors'
% 		set key spacing 1.5
	\begin{gnuplot}%[terminal=epslatex]
		set term epslatex size 3,2 
    set xlabel '\f{|\bar u_2|} [mm]' offset 1,1.1,0
		set ylabel '\f{R} [N/mm]' offset 1.5, 0
		set xtics (0, 0.6)
		set ytics (0,40)
		set nokey
		set xrange [0:0.6]
		set yrange [0:40]
    plot 'Figs/Data/Pbeam2da.sum' u ($1):(-1*$2) w lp 
	\end{gnuplot}
	\hspace{-5mm}
%     plot 'Figs/Data/20241130_08h37_light_alp0_5redIntHM2_8_2_10mds/feap/histVars.dat' u 5:($3*1000) w l title 'FEM \f{B_1}-\f{H_1}', 'Figs/Data/20241130_08h37_light_alp0_5redIntHM2_8_2_10mds/exe/histVarsRed.dat' u 4:($2*1000) w l title '10/2 IPs \f{B_1}-\f{H_1}', 'Figs/Data/20241130_08h37_light_alp0_5redIntHM2_8_2_10mds/feap/histVars.dat' u 6:($4*1000) w l title 'FEM \f{B_2}-\f{H_2}', 'Figs/Data/20241130_08h37_light_alp0_5redIntHM2_8_2_10mds/exe/histVarsRed.dat' u 5:($3*1000) w l title '10/2 IPs \f{B_2}-\f{H_2}'
\caption{Left: Plane-strain two-scale simulation of a beam with dimensions 10\f{\times}2\,mm, being clamped on the left end. Right: Reaction force over displacement at the right end.}
\label{beam}
\end{figure}

Figure~\ref{beam} (left) shows an exemplary two-scale simulation of a beam, being loaded on its right end by a prescribed macroscopic displacement \f{\bar u_2} in the vertical direction. The correspoding reaction force is depicted in the right part of the figure. For \f{p=10}, \f{\NipHR=8} and a macroscopic mesh consisting of 40\f{\times}20=800 elements, the total simulation time was \f{\sim}8.6\,s using 10 equal time steps. For comparison, a single scale simulation using the Ramberg-Osgood model with \f{p=10} is approximately 11 times faster, as it takes \f{\sim}0.8\,s. For a mesh with 160\f{\times}80=12800 elements, these figures increase to \f{\sim}146\,s and \f{\sim}20\,s, respectively. In this case, the single-scale simulation is only 7.3 faster than the two-scale model. A possible explanation for the different factors is that for gradual mesh refinement, the time required to solve the global linear equation systems grows disproportionally with respect to the numbers of degrees of freedom, in contrast to the effort due to the material law evaluation. Based on this hypothesis, the aforementioned factors could further be decreased by parallelization of the assembly process, which is beyond the scope of this work.
% 19.63
% 146 s
\subsection{Composite with large fibre diameter}
\begin{figure}[h]
\centering
	\hspace{-15mm}
% 		set arrow from 6,1 to 16,1 nohead dt 2
  \psfrag{a}{(a)}
%   \psfrag{}{}
  \includegraphics[width=0.3\textwidth]{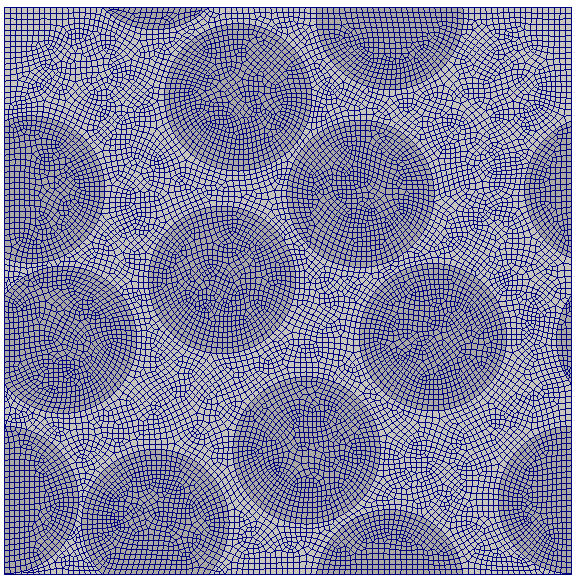}
	\hspace{-15mm}	
	\begin{gnuplot}%[terminal=epslatex]
		set term epslatex size 3.5,2.5 
		set title '(b) Max./avg. errors'
		set logscale x
    set xlabel '\f{\NipHR}' offset 1,.5,0
		set ylabel 'error \f{E} [\%]' offset .5, 0
		set xtics (5, 6, 7, 15, 18,  30)
		set ytics (1, 5, 10)
		set key spacing 1.5
		set key at 24.0, 12.5
		set xrange [4.8:30.2]
		set yrange [0:15]
		set arrow from 4.8,1 to 30.2,1 nohead dt 2
    plot 'Figs/Data/unseendata_errors_R_n5.dat' u 1:2 w lp notitle lc rgb 'red' pt 1 dt 2 lw 3, 'Figs/Data/unseendata_errors_R_n5.dat' u 1:3 w lp title '\f{p=5}' lc rgb 'red' pt 1 dt 2 lw 3, 'Figs/Data/unseendata_errors_R_n20.dat' u 1:2 w lp title '\f{p=20}' lc rgb 'dark-green' pt 2  dt 4 lw 3, 'Figs/Data/unseendata_errors_R_n20.dat' u 1:3 w lp notitle lc rgb 'dark-green' pt 2 dt 4 lw 3
	\end{gnuplot}
	\hspace{-5mm}
%     plot 'Figs/Data/20241130_08h37_light_alp0_5redIntHM2_8_2_10mds/feap/histVars.dat' u 5:($3*1000) w l title 'FEM \f{B_1}-\f{H_1}', 'Figs/Data/20241130_08h37_light_alp0_5redIntHM2_8_2_10mds/exe/histVarsRed.dat' u 4:($2*1000) w l title '10/2 IPs \f{B_1}-\f{H_1}', 'Figs/Data/20241130_08h37_light_alp0_5redIntHM2_8_2_10mds/feap/histVars.dat' u 6:($4*1000) w l title 'FEM \f{B_2}-\f{H_2}', 'Figs/Data/20241130_08h37_light_alp0_5redIntHM2_8_2_10mds/exe/histVarsRed.dat' u 5:($3*1000) w l title '10/2 IPs \f{B_2}-\f{H_2}'
\caption{(a) Mesh of the reinforced composite. (b) Maximum and average hyper-reduction errors \f{E} (see Eq.~\eqref{errorE}) over total number of integration points \f{\NipHR}. The linear-elastic fibres are represented by a single integration point in all cases except \f{p=20} and \f{\NipHR=30}, where three integration points were used.}
\label{errorsR}
\end{figure}

Next, the fibre diameter is increased (see Fig.~\ref{errorsR}, left), provoking intense strain fluctuations in between the fibres. 
The fully integrated reduced-order model is capable to accurately predict the finite element results with 11-14 modes, with errors typically in the order of 1\% and below. For \f{p=5}, the dependence of the hyper-reduction error on the number of integration points is comparable to the case with smaller fibre diameter (see red lines in Fig.~\ref{errorsR}, right). However, for \f{p=20}, in total 30 integration points are required to reach errors of \f{\sim}1\% and below (green lines).
% \\
% p=5, R, 4+1 IPs, 11 mds; doE3Ctraining: 2 min 30s; retraining: 28s\\
% p=5, R, 5+1 IPs, 11 mds; doE3Ctraining: 2 min 55s; retraining: 36s\\
% p=5, R, 6+1 IPs, 11 mds; doE3Ctraining: 3 min 29 s; retraining: 39s + 44s + 46s \\
% p=5, R, 8+1 IPs, 11 mds; doE3Ctraining: 3 min 57s; retraining: 44s\\
% p=20, R, 17+1 IPs, 14 mds; doE3Ctraining: 4 min 54 s; retraining: 55 s\\

\section{Summary and conclusion}
The hyper-reduction technique 'Empirically Corrected Cluster Cubature (E3C)', which was recently proposed in the context of magnetostatics, has been adopted to nonlinear mechanical computational homogenization. The number of integration points needed to reach hyper-integration errors in the order of 1\% depends on the microstructure and material nonlinearity, but has been found to be roughly similar to the number of modes. As the number of integration points approaches the theoretical limit\footnote{If the number of integration points falls below the limit of \f{\Nmd/n_{\rm T}}, the underlying equation systems in \f{\und{\xi}} become singular (with \f{n_{\rm T}} being the dimension of \f{\und{\hat \sigma}} and \f{\und{\hat \eps}}).}, it seems reasonable to focus next on reducing the number of modes (rather than integration points) to further improve performance. One approach to achieve this is problem-adopted training\citep{lange2024monolithic}. However, the origin of the high number of modes required in certain cases (low hardening, strong fluctuations) still needs to be better understood.
\begin{center}%\newpage
{\bf Acknowledgements} 
\end{center} 
This work was supported by the German Research Foundation (Deutsche Forschungsgemeinschaft, DFG) through the project A10 of the Collaborative Research Center SFB 1261 under Grant 286471992. This support is gratefully acknowledged.
% This research is associated to project 286471992 (Collaborative Research Center 1261), which is supported by the German Research Foundation (Deutsche Forschungsgemeinschaft, DFG). This support is gratefully acknowledged. 
% % This research was supported by the Deutsche Forschungsgemeinschaft (DFG, German Research Foundation) under project 286471992 (Collaborative Research Center 1261). 
% % Stephan Wulfinghoff and Christian Dorn would like to greatfully acknowledge this financial support.% of the German Science Foundation (DFG) related to the project '....
\begin{appendix}
\section{Preservation of the cluster average}\label{app1}
The proof of Eq.~\eqref{clustavg} reads
\begin{equation}
 \feps^q = \bH + \sum\limits_{k=1}^{\Nmd} \xi_k \tcH^q_k 
 \stackrel{\eqref{Htldq}}{=} \bH +\frac{1}{\Omega^q}\sum\limits_{p\in \cC^q} \sum\limits_{k=1}^{\Nmd} \xi_k \tcH_k (\fx^p) \OmegaFE^p = \frac{1}{\Omega^q}\sum\limits_{p\in \cC^q} \feps (\fx^p) \OmegaFE^p. 
\end{equation}
\section{Ramberg-Osgood law}\label{appRO}
The Ramberg-Osgood law reads
\begin{equation}
 \feps=\frac{1}{9\kappa}\tr{\fsigma}\fI + \frac{\fsigma'}{2\mu} + \frac{3}{2}\eps_0\left( \frac{\seq}{\sigma_0} \right)^p \frac{\fsigma'}{\seq}\label{RO}
\end{equation}
with the stress deviator~\f{\fsigma'}, von-Mises stress~\f{\seq=\sqrt{3/2}\,\|\fsigma'\|} and material parameters in Tab.~\ref{matPars} (\f{\kappa} and~\f{\mu} denote the bulk and shear modulus, respectively). 
\begin{table}[h]
  \centering
  \begin{tabular}{||c|c|c|c|c||}
  \hline
  \f{E_{\rm M}} [GPa] & \f{\nu_{\rm M}} & \f{\sigma_0} [MPa] & \f{\eps_0} [-]& \f{p}  \\ \hline
   210 & 0.3 & 200 &0.001 & see text \\
  \hline
  \end{tabular}
  \caption{Parameters of Ramberg-Osgood matrix ('M').}
  \label{matPars}
\end{table}
\end{appendix}
\bibliographystyle{elsarticle-harv}
\bibliography{lit}

\end{document}

%% file: formel_preprint.tex
\newcommand{\eps}{\varepsilon}
\renewcommand{\d}{{\,\rm  d}}

\newcommand{\T}[1]{{#1}^{\sf  T}}

\newcommand{\grad}[1]{{\rm grad}\left( #1 \right)}

\renewcommand{\div}[1]{{\rm div }\left( #1 \right)}

\newcommand{\tr}[1]{{\rm tr  }( #1 )}

\newcommand{\cblue }{\color{blue}}

\newcommand{\soutaa}[1]{}
\newcommand{\fempty}[1]{{}}
\newcommand{\f}[1]{\mbox{$ #1 $}}

\newcommand{\sty}[1]{\mbox{\boldmath $#1$}}

\newcommand{\styy}[1]{{\mathbb{#1}}}

\newcommand{\fu}{\sty{ u}}

\newcommand{\fx}{\sty{ x}}

\newcommand{\fzero}{\sty{ 0}}

\newcommand{\fI}{\sty{ I}}

\newcommand{\fN}{\sty{ N}}

\newcommand{\fU}{\sty{ U}}

\newcommand{\ffR}{\styy{ R}}

\newcommand{\fsigma}{\mbox{\boldmath $\sigma$}}

\newcommand{\feps}{\mbox{\boldmath $\varepsilon $}}
\newcommand{\fEps}{\mbox{\boldmath $\mathcal{E} $}}

\newcommand{\cC}{{\cal C}}

\definecolor{boxrahmen}{gray}{0.0}
\definecolor{boxhintergrund}{gray}{0.999}
\newsavebox{\tmpbox}